\documentclass[a4paper,fleqn]{cas-dc}
\usepackage[T2A]{fontenc}
\usepackage[utf8]{inputenc}
\usepackage[main=english, russian]{babel}
\usepackage[numbers]{natbib}

\usepackage{siunitx}
\usepackage{xcolor}
\usepackage{graphicx}
    \graphicspath{{figures/}}
\usepackage{tabularx}
\usepackage{multirow}

\begin{document}
\let\WriteBookmarks\relax
\def\floatpagepagefraction{1}
\def\textpagefraction{.001}
\shorttitle{Anechoic Chamber at ITMO University}
\shortauthors{F. Bikmukhametov et~al.}

\title[mode = title]{Applicability of Radiowave Anechoic Chambers for Acoustic Free-Field Measurements on the Example of the Chamber at ITMO University}

\author[1,2]{Farid Bikmukhametov}[orcid=0009-0004-6950-9216]
\ead{f.bikmukhametov@metalab.ifmo.ru}


\affiliation[1]{organization={School of Physics and Engineering, ITMO University},
                city={Saint Petersburg},
                postcode={197101},
                country={Russia}}
\affiliation[2]{organization={Laboratory ``Acoustics of Halls'', Research Institute of Building Physics},
                city={Moscow},
                postcode={127238},
                country={Russia}}

\affiliation[3]{organization={Soundproofing European Technologies LLC},
                city={Saint Petersburg},
                postcode={195027},
                country={Russia}}

\affiliation[4]{organization={Department of Architectural and Structural Design and Environmental Physics, Moscow State University of Civil Engineering},
                city={Moscow},
                postcode={129337},
                country={Russia}}

\author[3]{Ksenia Razrezova}[]


\author[3]{Roman Smolnitsky}[]


\author[3]{Yuri Shchelokov}[]


\author[4]{Nikolay Kanev}[orcid=0000-0001-6512-7131]


\author[5,1]{Mariia Krasikova}[orcid=0000-0002-5950-4807]
\ead{mariia.krasikova@gmail.com}
\affiliation[5]{organization={Devolm LLC},
                city={Krasnoyarsk},
                postcode={660049},
                country={Russia}}
                

\begin{abstract}
    Acoustic anechoic chambers allow free-field measurements required for the verification of effects under investigation and for the characterization of developing devices. However, the construction of an anechoic chamber is a labor-intensive process that requires a lot of resources, which is why these facilities are rather rare. An analogous statement can be made about radiowave chambers. At the same time, it is known that materials for radiowave absorption might absorb acoustic waves as well, and it can be expected that some chambers can be utilized for both electromagnetic and acoustic free-field measurements. This work examines the feasibility of the radiowave anechoic chamber of ITMO University (Saint-Petersburg, Russia) for acoustic measurements. The acoustic properties of the chamber coatings are estimated via measurements and numerical calculations. Characterization of sound pressure level, background noise level, and reverberation time is performed in accordance with the ISO 3745:2012 standard. The key conclusion is that the chamber can be considered as an acoustic anechoic chamber, but only for specific frequency ranges and distances from the source, which depend on the measurement directions.
\end{abstract}




\maketitle

\section{Introduction}

An essential stage of research and development work is the characterization of the systems under development, which typically requires high-precision measurements. Many of the testing procedures must be conducted under the free-field condition, when the amplitude is inversely proportional to the distance from the source~\cite{muller2013handbook}. When this condition is satisfied, it can be assumed that waves reflected from the surfaces of the surrounding space do not affect the results of measurements. For example, free-field conditions are required to determine the directivity and amplitude-frequency characteristics of loudspeakers~\cite{beranek2019acoustics}, perform microphone calibration~\cite{burnett1987freefield,zuckerwar2006calibration}, characterize musical instruments~\cite{rossing2014springer}, test hearing aids~\cite{iec_60118-0_2022} or investigate noise barriers~\cite{piechowicz2023sound, laxmi2022evaluation, ramadan2024studying, barros2024noise}.

Free-field conditions can be implemented for measurements conducted in anechoic chambers representing rooms in which the scattering is suppressed by using absorbing coatings covering the surfaces of the room~\cite{xu2019anechoic}. Depending on the locations of the absorbing coatings, chambers can be classified into two types: semi-anechoic and fully anechoic. In the first case, the coatings cover the walls and ceiling of the chamber, while in the second case the floor is also covered to minimize the influence of reflections on the measurements. Usually, coatings are represented by arrays of wedges made of porous materials, such as glass wool~\cite{velizhanina1957}, fiberglass~\cite{dombrovski1962, dombrovski1966, yuksek2022improvement, zhang2023review}, ultra-fine basalt fiber~\cite{makashov2017}, or phenolic felt~\cite{dombrovski1967}. At the same time, despite the abundance of various sound-absorbing materials, the quest to find the most efficient one is still open~\cite{efimtsov2001sound,bobrovnitskii2007impedance,christian2022absorber,haasjes2025solution,di2025improvement}.

\begin{figure*}[htbp!]
    \centering
    \includegraphics[width=0.9\linewidth]{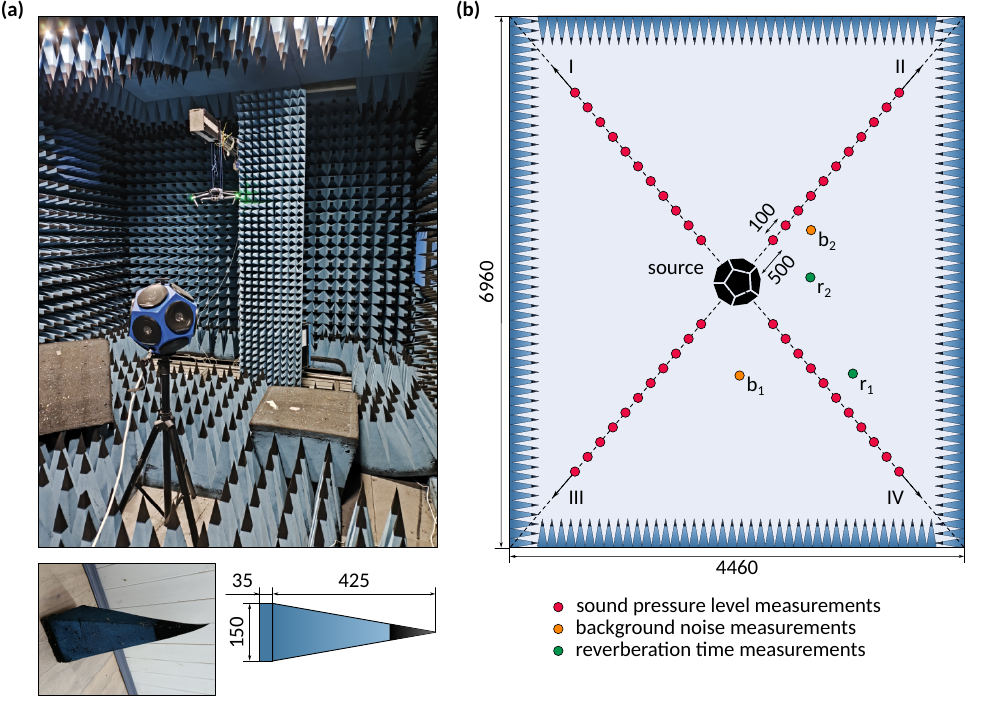}
    \caption{\textbf{Considered system.} (а) Photo of the anechoic chamber as well as photo and the schematic image of a pyramid element of the absorbing coating. (b) Measurement scheme with marked points at which the sound pressure level (red dots), background noise (orange dots) and reverberation time (green dots) are measured. The measurement directions are marked by indexes I -- IV.}
    \label{fig:geometry}
\end{figure*}

While anechoic chambers are crucially important facilities for acoustic measurement and testing, each camera represents a rather unique object~\cite{rivin1961,dombrovski1962,dombrovski1967,kopiev2017construction,vinogradov2019}, as the construction of the chambers is a laborious and resource intensive process. At the same time, anechoic chambers are constructed not only for acoustic measurements but also for electromagnetic ones~\cite{xu2019anechoic}. The coatings absorbing radio-waves can be made of carbon-impregnated polyurethane~\cite{khalid2017feasibility}, ferrite ~\cite{ellam1994update,naito1994anechoic}, foil-coated polymers~\cite{boiprav2024lightweight}, or activated charcoal made of coconut shells~\cite{rabelsa2019pyramidal}. It is noteworthy that some materials can absorb both acoustic and radio-waves. These might be carbon fibers~\cite{shen2014sound}, polymer nanocomposites~\cite{navidfar2022fabrication}, charcoal~\cite{sakthivel2021sound,khrystoslavenko2023sound}, or metamaterials~\cite{qu2022microwavea,gao2025broadband}. Therefore,it can be expected that in some spectral ranges radio-wave anechoic chambers can be considered as acoustic ones and vice versa.

The aim of the work is to determine the suitability of the radio-wave anechoic chamber of ITMO University (Saint-Petersburg, Russia) for free-field acoustic measurements. Characterization of the chamber is performed in accordance with the ISO 3745:2012 / GOST-ISO 3745-2014 standard, which defines the requirements imposed on the sound pressure levels and background noise level. Additionally, acoustic absorption coefficient of the chamber coatings is estimated. It is demonstrated that the chamber can be utilized as an acoustic anechoic chamber in the case of measurements near the sound source (at the distances up to $\SI{0.9}{\meter}$) at the frequencies up to $\SI{3150}{\hertz}$. The free-field condition is also satisfied for wider spectral ranges and at larger distances from the source, but whether the criteria for acoustic chambers are satisfied depends on the measurement directions. Therefore, the main conclusion of the work is that the radio-wave anechoic chamber of ITMO University is suitable for acoustic measurements under the free-field conditions at specific spectral ranges.

\section{Description of the chamber}
The chamber represents a room having the shape of a rectangular parallelepiped. Without coatings, the length of the walls is $4.46$ and $\SI{6.96}{\meter}$, and the height of the ceiling is$\SI{3.3}{\meter}$. The coatings represent arrays of pyramid elements with a square base [see Fig.~\ref{fig:geometry}(a)], which are made of open cell foam rubber saturated with finely-dispersed carbon particles with admixture of metal oxides (about $2$ -- $3$\%). The height of each pyramid is $\SI{425}{\milli\meter}$, while the thickness of the base is $\SI{35}{\milli\meter}$, and the width of the base is $\SI{150}{\milli\meter}$. To improve the radio-wave absorption, the apex parts are treated with a special solution the composition of which is unknown. It should be noted that there is a recess in the floor of the chamber that is $\SI{0.15}{\meter}$ deep, $\SI{1.3}{\meter}$ wide and $\SI{3.55}{\meter}$ long. Due to operational necessity, the recess is not covered by the absorbing coatings, as can be seen in Fig.~\ref{fig:geometry}(a). In addition, some parts of the ceiling are covered with uniform layers instead of pyramid arrays.

\section{Methods}
The determination of the acoustic characteristics of the chamber is provided in accordance with ISO 3745:2012 / GOST-ISO 3745-2014~\cite{iso3745-2012, gost_iso3745_2015}. Pressure fields are generated by the omnidirectional sound source OED-SP360 with the amplifier OED-PA360. Measurements of spatial distributions of the sound pressure level (SPL) are performed with the sound level meter Bruel\&Kjaer type 2250 and the microphone Sennheiser e604. The background noise and reverberation time measurements are conducted in points $\text{b}_{1,2}$ and $\text{r}_{1,2}$, respectively, while the SPL measurements are carried out in several points located along $4$ trajectories directed from the source to the corners of the camera, as shown in Fig.~\ref{fig:geometry}(b). Specifically, each measurement direction contains $11$ points such that the closest to the source is located at the distance $\SI{500}{\milli\meter}$ from it, and the consequent points are arranged with the step $\SI{100}{\milli\meter}$. The measurement time at each point is $\SI{30}{\second}$.

Measurements of the absorption coefficient of the coating material are conducted in the impedance tube using the two-microphone method, in accordance with ISO 10534-2:2023~\cite{iso10534_2_2023}. The tube is characterized by a circular cross-section with radius $\SI{55}{\milli\meter}$. The sound source is placed at one end of the tube, whereas the sample is at the other. Acoustic pressure is measured by microphones located at the distances $730$ and $\SI{790}{\milli\meter}$ from the sample. Then, the transfer function is calculated to estimate the reflection coefficient $R$ and the absorption coefficient $A = 1 - R$. The spectral range within which the reliable results can be obtained is limited by the geometry of the tube as high frequencies are associated with the excitation of non-piston modes. Specifically, the upper frequency is set to be $\SI{1800}{\hertz}$.

\begin{figure*}[ht!]
    \centering
    \includegraphics[width=0.9\linewidth]{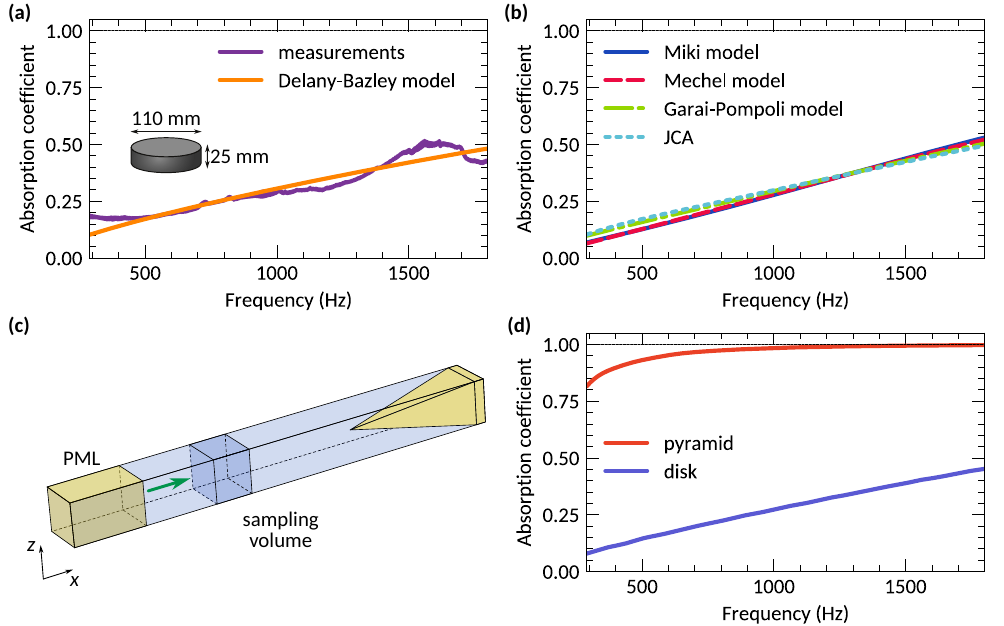}
    \caption{\textbf{Acoustic properties of the coatings.} (а) Measured absorption coefficient of the coating material and its approximation via Delany-Bazley model. The sample represents a disk with the radius $\SI{55}{\milli\meter}$ and the thickness $\SI{25}{\milli\meter}$. (b) Approximation of the absorption coefficient using different empirical models. (c) Schematic illustration of the numerical model for calculation of absorption coefficient spectra of pyramid elements. Floquet periodicity conditions are imposed on the boundaries of the computational domain lying in $xy$ and $xz$ planes. (d) Numerically calculated absorption coefficient for the disk and pyramid samples characterized by the material with the flow resistivity $\SI{3437}{\pascal.\second/\meter\squared}$ in the framework of the Delany-Bazley model.}
    \label{fig:pyramids}
\end{figure*}

Numerical simulations are conducted with the help of COMSOL Multiphysics using the ``Pressure Acoustics, Frequency Domain'' module. The incident field is introduced as a plane wave with the amplitude $\SI{0.1}{\pascal}$ using the ``Background Pressure Field'' feature. The absorption coefficient is calculated as $A = 1 - |r|^2$, where $r$ is the reflection coefficient defined as
\begin{equation}
    r = \frac{1}{V} \int \frac{p_s}{p_b} ~~ \mathrm{d}V,
\end{equation}
where $p_s$ is the pressure of the scattered field, $p_b$ is the pressure of the background (incident) field, and the integration is performed over a small closed volume $V$ [see. Fig~\ref{fig:pyramids}(c)].

\section{Results}
\subsection{Absorbing Coatings}
Due to limitations imposed by the dimensions of the measurement setup, the acoustic characteristics of the pyramid elements could not be estimated experimentally. Instead, the absorption coefficient of the coating material is measured for the disk-like sample cut from a pyramid. The radius of the sample is $\SI{55}{\milli\meter}$, and the thickness is $\SI{25}{\milli\meter}$. According to the results of measurements shown in Fig.~\ref{fig:pyramids}(а), the absorption coefficient takes values from $0.17$ to $0.5$ within the spectral range $290$ -- $\SI{1800}{\hertz}$.

Since the sample is made of a porous material, its acoustic characteristics can be described using conventional empirical models in which the impedance and wavenumber are defined using flow resistivity $\sigma$. In particular, the experimental date is approximated by estimating the appropriate value of $\sigma$ within the framework of the five most common models, such that the fitting accuracy is evaluated by calculating the standard deviation. The results of the estimations are listed in Tab~\ref{tab:empyric_models}. It can be stated that the Delany-Bazley model~\cite{delany1970acoustical} provides the most accurate fitting, as shown in Fig.~\ref{fig:pyramids}(a). An almost identical result is achieved using the Johnson-Champoux-Allard (JCA)~\cite{allard1992new} and Garai-Pompoli models~\cite{garai2005simple}. Meanwhile, the Miki~\cite{miki1990acoustical} and Mechel~\cite{mechel1986absorption} models provide a bit worse result, but the deviation is small, as follows from Fig.~\ref{fig:pyramids}(b).

\begin{table}[ht!]
    \renewcommand{\arraystretch}{1.2}
    \newcolumntype{Y}{>{\centering\arraybackslash}X}
    \centering
    \caption{Fitted values of flow resistivity of coating material for different empirical models.}
    \label{tab:empyric_models}
    \begin{tabularx}{0.8\linewidth}{>{\hsize=0.8\hsize}X|Y}
        Model & Flow resistivity (Pa$\cdot$s/m$^2$) \\
        \hline
        Delany-Bazley & 3437 \\
        Miki & 5994 \\
        Mechel & 5136 \\
        Garai-Pompoli & 6125 \\
        JCA & 8701 \\
    \end{tabularx}
\end{table}

Nevertheless, absorbing coatings are made of pyramid elements rather than uniform layers with a fixed thickness. At the same time, pyramid elements are known to have higher absorption compared to a non-structured layer of a porous material~\cite{cox2016acoustic}. The estimation of the absorption coefficient is performed using numerical calculations. Specifically, a waveguide with a square cross section is considered such that the pyramid is placed at one end, while the second end is supplemented by the perfectly matched layer (PML). To imitate an infinitely periodic (along $y$ and $z$ axes) structure, Floquet periodicity conditions are imposed on the boundaries of the computational domain that lie in the $xy$ and $xz$ planes. The porous material is described with the Delany-Basley model with flow resistivity $\SI{3437}{\pascal.\second/\meter\squared}$ defined in the fitting step. Calculations are performed for both the disk and the pyramid samples. Note that the influence of the solution with which the apex is treated is not taken into account, and the pyramid is considered to be uniform.

According to Fig.~\ref{fig:pyramids}(d), the results of calculations for the disk are expectedly close to the analytical approximation. The absorption coefficient in this case is below $0.5$ for the whole considered spectral range. The situation changes drastically for the pyramid element, such that at $\SI{290}{\hertz}$ the absorption coefficient is about $0.815$ and then increases to $0.996$ with increasing frequency. The spectral dependence in this case resembles a logarithmic function such that the absorption coefficient starts to overcome the value of $0.9$ at frequencies above $\SI{410}{\hertz}$.

Therefore, it can be concluded that in terms of acoustic characteristics, chamber coatings are characterized by a reasonably high absorption coefficient (within the considered spectral range $290$ -- $\SI{1800}{\hertz}$) comparable to the one of materials utilized for coatings of acoustic anechoic chambers~\cite{aso1966sound, feng2025latest,mohammadi2022mechanical, liu2025simulation, ikpekha2025effect}.

\subsection{Background noise}
One of the key characteristics of anechoic chambers is the background noise level, as it directly affects the results of conducted tests and measurements. According to ISO 3745:2012 / GOST-ISO 3745-2014, the requirements imposed on the background noise level are satisfied when the equivalent SPL of the background noise, averaged over the measurement points, is at least $\SI{10}{\decibel}$ lower than the corresponding equivalent SPL of the noise source for third-octave bands with center frequencies ranging from $250$ to $\SI{5000}{\hertz}$. 

\begin{figure}
    \centering
    \includegraphics[width=0.925\linewidth]{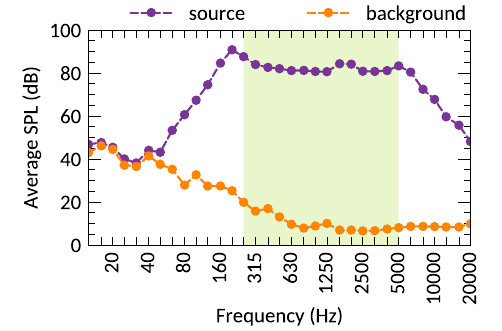}
    \caption{\textbf{Background noise level.} 
    Spectra of the averaged equivalent sound pressure levels (SPL) of background noise and the equivalent SPL of the utilized noise source. Measurements are carried out at the center frequencies of the third-octave bands. Shaded are indicates the spectral range from $250$ to $\SI{5000}{\hertz}$.}
    \label{fig:background_field}
\end{figure}

The comparison of the measured equivalent SPL of the background noise and the noise source is shown in Fig.~\ref{fig:background_field}. On average, the SPL of the background noise is $\SI{72}{\decibel}$ than SPL of the source, such that the minimal difference is $\SI{65.7}{\decibel}$ (within the range $250$ -- $\SI{5000}{\hertz}$). Hence, the requirements imposed on the background noise level are fully satisfied.

\subsection{Reverberation time}
Reverberation time defines the rate of SPL decrease after the switching off of the source. Therefore, the lower the reverberation time, the smaller influence the reflected fields have on the total field distribution in the chamber. For an ideal anechoic chamber the reverberation time is equal to $\SI{0}{\second}$.

\begin{figure}
    \centering
    \includegraphics[width=0.925\linewidth]{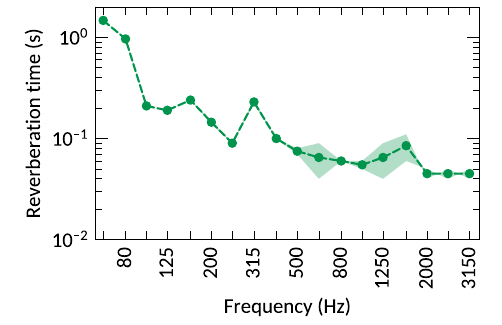}
    \caption{\textbf{Measured reverberation time.} The figure shows averaged reverberation time in the  $1/3$-octave frequency band. Shaded area indicates the spread of values.}
    \label{fig:reverberation}
\end{figure}

For the considered chamber, the reverberation time is the average of the reverberation times measured at two points marked in Fig.~\ref{fig:geometry}(b). The resulting spectrum (in the one-third octave bands) is shown in Fig.~\ref{fig:reverberation}. According to the obtained dependence, the reverberation time quickly drops from $\SI{1.5}{\second}$ at $\SI{63}{\hertz}$ to $\SI{0.21}{\second}$ at $\SI{100}{\hertz}$, and for the larger frequencies it remains in-between $0.05$ and $\SI{0.24}{\second}$. 

\subsection{Sound Pressure Levels}
\begin{figure*}[ht!]
    \centering
    \includegraphics[width=0.95\linewidth]{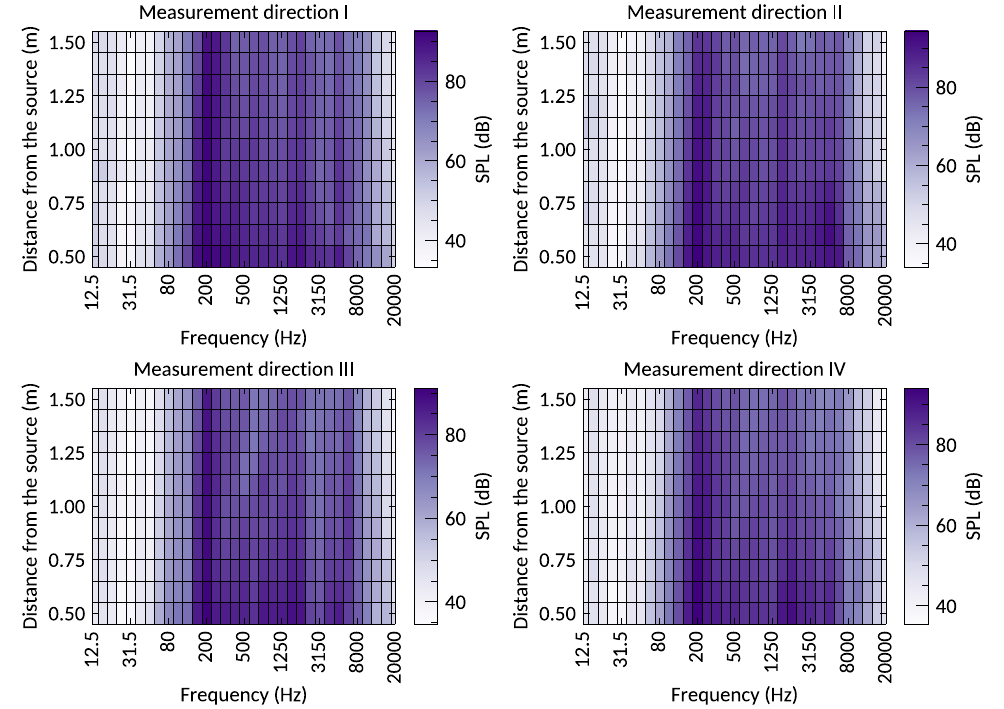}
    \caption{\textbf{Measured sound pressure levels.} Sound pressure level (SPL) spectra as functions of the distance from the source for different measurement directions.}
    \label{fig:SPL}
\end{figure*}

The SPL spectra measured for various distances from the noise source are shown in Fig.~\ref{fig:SPL}. Crucially, the SPL for different measurement directions deviate from each other, which is an indication of the field inhomogeneities caused by reflections from the chamber walls. Therefore, it might be expected that the criteria for the chamber's suitability for different measurement directions will be met for different spectral ranges and distances from the source. In addition, it should be mentioned that the inequivalence of measurement directions can be related to non-uniformities in the directivity pattern of the source, which was not verified before the measurements. However, the conclusions of the work are based on the assumption that the source is omnidirectional.

The compliance of the chamber with the criteria imposed on anechoic chambers is defined by the comparison of measured SPL with the corresponding theoretical values calculated as
\begin{equation}
    L_p(r_i) = 20 \lg\left(\frac{a}{r_i - r_0} \right),
\end{equation}
where $a$ is a parameter determined by the sound power of the noise source and $r_0$ is the displacement of the acoustic center along the trajectory of the microphone displacement from the origin (i.e. the center of the measurement spherical surface) from which the distance $r_i$ is counted. These parameters can be defined as
\begin{equation}
    a = \frac{M r_0^2 + \sum\limits_{i=1}^M r_i^2 - 2r_0 \sum\limits_{i=1}^M r_i}{\sum\limits_{i=1}^Mr_i q_i - r_0 \sum\limits_{i=1}^M q_i},
\end{equation}
\begin{equation}
    r_0 = \frac{\sum\limits_{i=1}^M r_i \sum\limits_{i=1}^M r_i q_i - \sum\limits_{i=1}^M r_i^2 \sum\limits_{i=1}^M q_i}{\sum\limits_{i=1}^M r_i \sum\limits_{i=1}^M q_i - M \sum\limits_{i=1}^M r_i q_i},
\end{equation}
where $q_i = 10^{-0.05 L_{p,i}}$. Here, $L_{p,i}$ is the SPL measured in the $i$-th point and $M$ is the total number of points along the measurement direction. Deviation of the measured SPL from theoretical predictions, defined as
\begin{equation}
    \Delta L_p = L_{p,i} - L_p(r_i),
\end{equation}
should not exceed the values listed in Tab.~\ref{tab:SPL_deviation_ref}.

\begin{table}[ht!]
    \renewcommand{\arraystretch}{1.2}
    \newcolumntype{Y}{>{\centering\arraybackslash}X}
    \centering
    \caption{The maximum acceptable deviations of measured sound pressure levels from the theoretical values.}
    \label{tab:SPL_deviation_ref}
    \begin{tabularx}{\linewidth}{Y>{\hsize=0.5\hsize}Y}
        Central frequency of the one-third octave band (Hz) & Acceptable deviations (dB) \\
        \hline
        $\leq 630$ & $\pm 1.5$ \\
        from $800$ to $5000$ & $\pm 1.0$ \\
        $\geq 6300$ & $\pm 1.5$ \\
    \end{tabularx}
\end{table}

\begin{figure*}[ht!]
    \centering
    \includegraphics[width=0.95\linewidth]{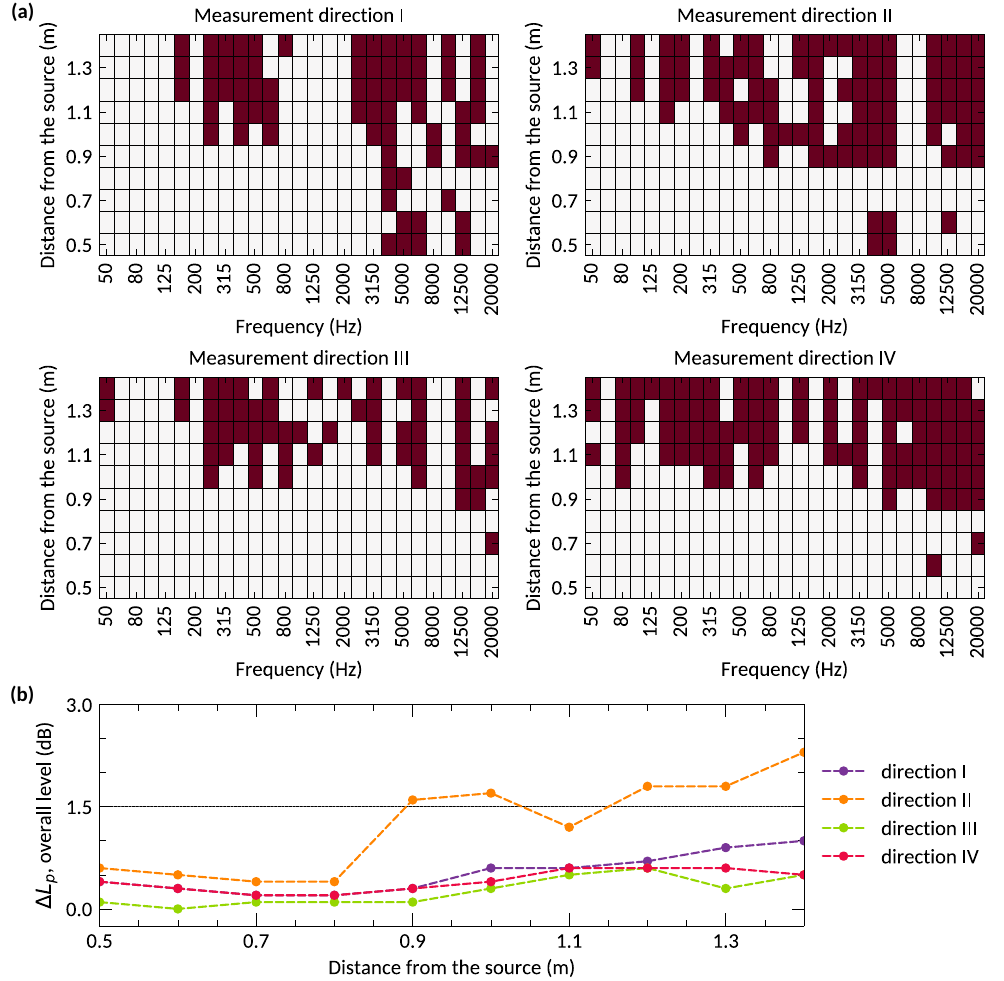}
    \caption{\textbf{Camera compliance with suitability criteria.} (a) Diagram indicating feasible deviation of the measured sound pressure level (SPL) from the corresponding theoretical predictions, $\Delta L_p$, for different measurement directions. Color indicates the frequencies in one-third octave band and distances from the source at which the values of $\Delta L_p$ exceed the feasible limits defined for anechoic chambers (in accordance with Tab.~\ref{tab:SPL_deviation_ref}).
    (b) Deviation of the measured overall SPL from the theoretical predictions.}
    \label{fig:SPL_deviation}
\end{figure*}

Figure~\ref{fig:SPL_deviation}(a) demonstrates spectral ranges (in the one-third octave band) and distances from the source for which the requirements for acoustic anechoic chambers are satisfied. To be specific, the cases where $\Delta L_p$ exceeds the feasible values are marked with red color on the diagram. Particular values are omitted for clarity, although it should be mentioned that the largest deviation reaches the value of $\SI{15.5}{\decibel}$. 
In general, the chamber meets the requirements only for a limited set of frequencies and distances to the source. One of the reasons behind this result might be the non-uniform distribution of the absorbing coating throughout the chamber, as mentioned in its description. At the same time, the maps for deviations $\Delta L_p$ are not equivalent to each other, which might be an indication of the presence of scattering. Nevertheless, when the distance from the source is $0.5$ -- $\SI{0.8}{\meter}$ the chamber can be considered as an anechoic in the spectral range of $50$ to $\SI{3150}{\hertz}$ regardless of the measurement direction. The spectral range is much larger for directions III and IV, where the upper frequency limits are $16000$ and $\SI{8000}{\hertz}$, respectively. Larger distances from the source to a measurement point imply larger deviations of the measured SPL from the corresponding theoretical values. It might be argued that the amplitudes of the reflected fields are much smaller than the amplitude of the generated field in the vicinity of the source. Hence, the distortions caused by the reflections are insignificant for the measurements near the source. For other distances, the contribution of the reflected fields becomes tangible, and consequently, the free-field conditions are satisfied only at some spectral ranges, which are listed in Tab.~\ref{tab:criteria}. It should be noted that when the overall SPL is considered, the deviation of the measured values from the theoretical ones does not exceed the level of $\pm \SI{1}{\decibel}$ for all measurement directions except the direction II [see Fig.~\ref{fig:SPL_deviation}(b)]. Furthermore, the values of $\Delta L_p$ are below $\SI{0.6}{\decibel}$ for all distances from the source up to $\SI{0.8}{\meter}$, which additionally indicates that the requirements for the anechoic chambers are satisfied. 

\begin{table}[ht!]
    \renewcommand{\arraystretch}{1.2}
    \newcolumntype{Y}{>{\centering\arraybackslash}X}
    \centering
    \caption{Frequency ranges and distances to the source for which the chamber can be considered anechoic.}
    \label{tab:criteria}
    \begin{tabularx}{0.9\linewidth}{YYY}
        Measurement direction  &  Frequency range (Hz)  &  Distance from the source (m) \\
        \hline
        \multirow{3}{*}{I}  &  50 -- 3150  &  0.5 -- 0.9 \\
          &  50 -- 125  &  1.0 -- 1.4 \\
          &  800 -- 2000 &  1.0 -- 1.4 \\
        \hline
        \multirow{3}{*}{II}  &  50 -- 1250  &  0.5 -- 0.9 \\
          &  1600 -- 10000  & 0.5 -- 0.8 \\
          &  63 -- 80  &  1.0 -- 1.4 \\
        \hline
        \multirow{3}{*}{III}  &  50 -- 10000  & 0.5 -- 0.9 \\
          & 63 -- 125  &  1.0 -- 1.4 \\
          & 1000 -- 2000  &  1.0 -- 1.3 \\
        \hline
        \multirow{3}{*}{IV}  &  50 -- 4000 &  0.5 -- 0.9 \\
          &  1600 -- 2500  &  1.0 -- 1.3
    \end{tabularx}
\end{table}

\section{Conclusion}
The radiowave anechoic chamber of ITMO University (Saint-Petersburg, Russia) is analyzed for suitability criteria imposed on acoustic anechoic chambers. The results of the characterization indicate that the camera can be utilized for free-field acoustic measurements, in agreement with the ISO 3745:2012 / GOST-ISO 3745-2014 standard. However, it should be highlighted that the requirements defined by the standard are satisfied only for specific distances from the noise source and at limited frequency ranges, which are listed in Tab.~\ref{tab:criteria}.

\section*{Author Contributions}
F.B. measured acoustic characteristics of absorbing coatings and processed the results. K.R., R.S., Y.S. and F.B. performed the measurements of sound pressure levels, background noise, and reverberation time of the chamber, K.R., R.S., and Y.S. processed the results. M.K. performed numerical calculations. M.K., Y.S. and N.K. supervised the work. All authors contributed to the analysis of the results and the preparation of the manuscript. 

\section*{Declaration of Competing Interest}
The authors declare that they have no known competing financial interests or personal relationships that could have appeared to influence the work reported in this paper.

\section*{Data availability}
The data that support the findings of this study are available from the corresponding author upon reasonable request.

\section*{Acknowledgements}
M.K. thanks Sergey Krasikov for useful discussions and comments.

\bibliographystyle{unsrt}
\bibliography{references.bib}

\end{document}